\documentclass{research4cacm}

\def\BibTeX{{\rm B\kern-.05em{\sc i\kern-.025em b}\kern-.08em
    T\kern-.1667em\lower.7ex\hbox{E}\kern-.125emX}}

\usepackage{url}
\usepackage{booktabs}

\usepackage[official]{eurosym}
\usepackage{subcaption}
\usepackage{multirow}
\usepackage{tikz}
\usetikzlibrary{arrows.meta,arrows}
\usepackage[noend]{algpseudocode}
\usepackage{algorithm}
\usepackage{ragged2e}
\usepackage{pgfplots}
\usepackage{xcolor}
\pgfplotsset{compat=1.16}

\definecolor{matlab_color_1}{RGB}{000, 114, 189}
\definecolor{matlab_color_2}{RGB}{217, 083, 025}
\definecolor{matlab_color_3}{RGB}{237, 177, 032}
\definecolor{matlab_color_4}{RGB}{126, 047, 142}
\definecolor{matlab_color_5}{RGB}{119, 172, 048}
\newcommand{\csvline}[2]{\addplot+[mark=none, line width=1pt, solid, thick, color=#1] table[col sep=comma]{#2};}
\newcommand{\csvdotted}[2]{\addplot+[mark=none, line width=1pt, dotted, ultra thick, color=#1] table[col sep=comma]{#2};}
\newcommand{\csvdashed}[2]{\addplot+[mark=none, line width=1pt, dashed, thick, color=#1] table[col sep=comma]{#2};}
\newcommand{\csvdenselydashed}[2]{\addplot+[mark=none, line width=1pt, densely dashed, thick, color=#1] table[col sep=comma]{#2};}
\newcommand{\csvdenselydotted}[2]{\addplot+[mark=none, line width=1pt, densely dotted, ultra thick, color=#1] table[col sep=comma]{#2};}
\newcommand{\csvpointstwo}[2]{\addplot+[mark=*, only marks, color=#1,mark options={solid,fill=#1,scale=0.5}] table[col sep=comma, x index={0}, y index={1}, y error index={2}]{#2};}

\makeatletter
\renewcommand{\secfnt}{\fontfamily{ptm}\fontsize{12}{14}\bfseries}
\renewcommand{\subsecfnt}{\fontfamily{ptm}\fontsize{11}{13}\itshape}
\def\section{%
    \@startsection{section}{1}{\z@}{-10\p@ \@plus -4\p@ \@minus -2\p@}% GM
    {14\p@}{\secfnt\@ucheadtrue}%
}

\def\subsection{%
    \@startsection{subsection}{2}{\z@}{-8\p@ \@plus -2\p@ \@minus -\p@}
    {14\p@}{\secfnt}%
}
\def\subsubsection{%
    \@startsection{subsubsection}{3}{\z@}{-8\p@ \@plus -2\p@ \@minus -\p@}%
    {14\p@}{\subsecfnt}%
}
\expandafter\def\expandafter\abstract\expandafter{\abstract\vspace{-\baselineskip}}
\makeatother

\makeatletter
\def\@copyrightspace{\relax}
\makeatother

\begin{document}

\title{Graph Theory for Consent Management: A New Approach for Complex Data Flows
\thanks{\copyright Copyright held by the owner/author(s). This is a minor revision of the paper entitled ``Consent Management in Data Workflows: A Graph Problem'' that was published in Proceedings of the 26th International Conference on Extending Database Technology (EDBT), March 28-31, 2023 on OpenProceedings.org. DOI: http://dx.doi.org/10.48786/edbt.2023.61.}
}

\numberofauthors{3} 
\author{
% 1st. author
\alignauthor
Dorota Filipczuk\titlenote{This research was conducted while the author was at the University of Southampton.}\\
       \affaddr{Microsoft}\\
       \affaddr{Dronning Eufemias gate 71}\\
       \affaddr{0194 Oslo, Norway}\\
       \email{dorotaf@acm.org}
% 2nd. author
\alignauthor
Enrico H. Gerding\\
       \affaddr{University of Southampton}\\
       \affaddr{University Road}\\
       \affaddr{Southampton SO17 1BJ, UK}\\
       \email{eg@ecs.soton.ac.uk}
% 3rd. author
\alignauthor 
George Konstantinidis\\
       \affaddr{University of Southampton}\\
       \affaddr{University Road}\\
       \affaddr{Southampton SO17 1BJ, UK}\\
       \email{G.Konstantinidis@soton.ac.uk}
}

\maketitle
\begin{abstract}
Through legislation and technical advances users gain more control over how their data is processed, and they expect online services to respect their privacy choices and preferences. However, data may be processed for many different purposes by several layers of algorithms that create complex data workflows. To date, there is no existing approach to automatically satisfy fine-grained privacy constraints of a user in a way which optimises the service provider's gains from processing. In this article, we propose a solution to this problem by modelling a data flow as a graph. User constraints and processing purposes are pairs of vertices which need to be disconnected in this graph. In general, this problem is NP-hard, thus, we propose several heuristics and algorithms. We discuss the optimality versus efficiency of our algorithms and evaluate them using synthetically generated data. On the practical side, our algorithms can provide nearly optimal solutions for tens of constraints and graphs of thousands of nodes, in a few seconds.
\end{abstract}

\section{Introduction}
Personal data processing is at the core of many computing systems. In some complex ones, such as those owned by Netflix, Meta or Amazon, users' data is automatically processed by microservices. Typically, personal data enters the system through front-end applications, which send requests to a subset of services. There, each individual service performs a specific business function independent from the rest of the services, often producing predictions or inferences that are in turn consumed by other services. For instance, in a social media system, the user's location may be used by a service responsible for processing user profile information that then sends this data to a usage analytics service and to a service for suggesting groups to join. Consequently, the inferences made by the usage analytics service may be used by an ads-ranking service and the predicted groups of user's interest may be used by a service recommending products to buy. These services next call other services, creating data flows that span over multiple layers of computation. The ultimate goal of this data processing is to eventually satisfy certain business goals by which the service provider gains utility e.g., the ads ranking service is used to increase revenue through personalised advertisement and the product recommendation service may serve the purpose of collecting commission on product sale.

However, individuals should have control over how their data is used. Particularly, in certain regulatory frameworks such as the GDPR \cite{gdpr}, unless there exists another legal basis, user's consent is necessary to legally allow personal data to be processed. By opting out of data processing, e.g., not allowing their location data to be used for personal advertisement or product recommendation, users put constraints on the data flow. Enforcing these constraints affects the utility of the service provider who would like the utility loss minimised. What complicates the task even more is its large scale: in modern computing systems, there may be many stages of data processing, in a data flow that involves hundreds of nodes. In fact, Meta's microservice topology contains over 12 million service instances and over 180,000 communication edges between services \cite{huye2023lifting}. In such cases, stopping the data flow to the ads-ranking service or the product recommendation service may affect the quality of the inferences that other services use, and thus, the utility of the service provider. 

In this article, we propose a novel approach to finding optimal ways of satisfying privacy constraints automatically while minimising the utility loss for the service provider.
We model the data flow as a graph and privacy constraints as pairs of vertices in the graph. Our problem definition is generic, allowing the utility of a purpose node to be a black-box function of the subgraph connected to that node. 
We formulate the problem as an optimisation problem, where pairs of graph vertices must be disconnected such that utility is maximised (Section~\ref{Section: Consented Data Workflow Problem}) and demonstrate it on an example use case (Section~\ref{Section: Example}). We then present a natural instantiation of the generic problem where the utility functions are linearly additive (Section~\ref{Section: Additive Model}).

We present five generic heuristics for implementing the privacy constraints into data workflows (Section~\ref{Section: Algorithms}) with an optimal or approximately optimal global utility, depending on the actual utility function used. We further analyse the (non-)optimality of our heuristics in Section~\ref{Section: Properties}. For our experiments (Section~\ref{Section: Evaluation}) we implement the aforementioned linearly additive instance of the problem.

Notably, we show that, although computing the optimal solution can be very time-consuming, the proposed heuristics can provide good and efficient approximation alternatives. Our approach opens a range of new problems that we discuss in Section~\ref{section:openproblems}. Finally we juxtapose our approach with related work in Section~\ref{Section: Related Work}.

\section{Problem Formulation}
\label{Section: Consented Data Workflow Problem}

We focus on what we call the \emph{Consented Data Workflow} problem (\textsc{CDW}): given user constraints expressed in terms of the vertices that they do not wish to be connected, find a subgraph of the original workflow where these constraints are satisfied. In the face of alternative solutions, the optimal one should minimise the utility loss for the service provider from applying the constraints. In other words, we are looking for the utility-maximising solution subject to the users' privacy constraints.

Formally, our data processing model is a directed graph $G = (V, E)$ with a set of edges $E$ representing the data flow and a set of vertices $V$ representing the stages of data processing. We distinguish three kinds of vertices, i.e. $V = V^U \cup V^A \cup V^P$, where $V^U$ is a set of user data vertices, which represent the types of data collected directly from the user, $V^A$ is a set of algorithm vertices, which represent data processing algorithms that take one or more data types as input, and $V^P$ is a set of purpose vertices, which represent the end goals of data processing.

Furthermore, satisfying the given purposes is what brings service providers utility. For any vertex in $V^P$, we will have an associated utility that reflects the value processing that this purpose brings to service providers. In practice, the service providers' valuation depends on factors such as the accuracy of the datasets used as the input to the processing algorithms \cite{li2014theory,ghorbani2019data}. In particular, where data processing is a multi-stage process, the utility is affected by all stages and all datasets processed for the purpose. 

Therefore, in order to calculate the utility of data processing for a given purpose, in our model we look for all vertices and edges that carry the data workflow to the given purpose vertex. Formally, we say that a vertex $v_i \in V$ is \textit{reachable} from a vertex $v_j \in V$ if there exists a path in $G$, $\{(v_1, v_2)$,$(v_2,v_3)$,$\dots$ $(v_{k-1},v_k)\}$  such that $v_1 = v_j$ and $v_k = v_i$. For each purpose vertex $p \in V^P$ in our graph $G$, the \textit{reachability subgraph} of $p$ is the graph $G_p = (V_p, E_p)$ where $V_p \subseteq V$ is the set of the vertices that $p$ is reachable from and $E_p \subseteq E$ is the set of edges from $G$ that connect them.  If an edge is removed from $G$, the reachability subgraph of one or more purpose vertices is affected. In general, we write $\mathcal{R}(G_p)$ to denote the set of all subgraphs of the reachability subgraph of $p$ in $G$ - that are still reachability graphs when some edges are removed. Then, to calculate the utility of fulfilling a purpose, for each purpose vertex $p \in V^P$ we define a utility function $u_p : \mathcal{R}(G_p) \rightarrow \mathbb{R}_0^+$, which is a function of a reachability subgraph of $p$.

While $u_p$ can be an arbitrary function dependent on the valuation (or more general, the importance) of datasets in the corresponding reachability subgraph, the valuations of some datasets may influence the valuations of others.  To describe the relationships between these valuations in our model, we define a valuation function $\pi : E \rightarrow \mathbb{R}_0^+$, representing the valuation of the data propagating through the edge in the data processing system. As we later describe in Section~\ref{Section: Additive Model}, given the reachability subgraph $G_p = (V_p, E_p)$ of a vertex $p \in V^P$, the utility function $u_p(G_p)$ at $p$ can be defined as a function of the valuations of edges in $E_p$.

In our setting, a user constraint denotes their choice to opt out their data, represented by a user vertex in $V^U$, from being used for a purpose vertex in $V^P$.

Formally, our set of constraints is a set $\mathcal{N} = \{(v_s, v_t) \: | \: v_s \in V^U, v_t \in V^P\}$. In order to satisfy the constraints, the initial graph $G$ needs to be modified by removing one or more edges that belong to the paths between pairs $(v_s, v_t)$, such that the utilities $u_p$ are maximised. In essence, our problem is a multi-objective optimisation problem, where the objectives are to maximise $u_p$, for all $p \in V^P$. The most common approach to multi-objective optimization is to turn the problem into a single-objective optimization using a weighted sum \cite{marler2004survey}. This allows us to define the utility of $G$ as:

\begin{equation}
\label{eq:global-utility-function}
U(G) = \sum_{p \in V^P} w_p u_p(G_p),
\end{equation}

where $w_p$ is the weight of the purpose corresponding to vertex $p$ and $G_p$ is the reachability subgraph of $p$. Therefore, given $\mathcal{N}$, the \textsc{CDW} problem is to find the consented subgraph of $G$:
\begin{equation}
\label{eq:optimisation_problem}
    G^{*} = arg \: \max_{G'} \: U(G')
\end{equation}
where $G' = (V, E')$ is a subgraph of $G$, $E' \subseteq E$, and there is no path from $s$ to $t$ for each $(s, t) \in \mathcal{N}$.

\section{Recommend Product Data Flow}
\label{Section: Example}

\begin{figure}
\includegraphics[width=9cm]{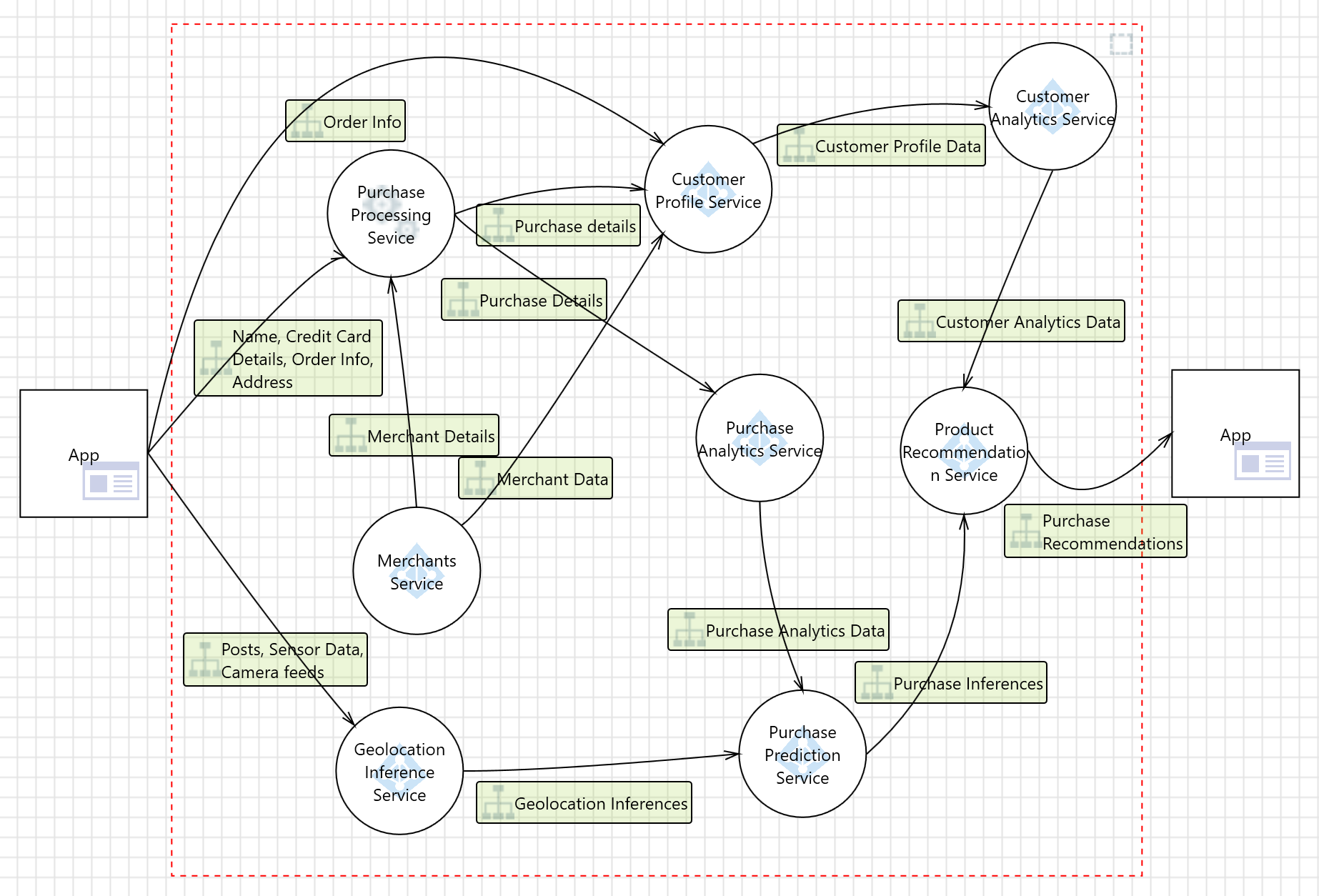}
\caption{A DFD created with Microsoft Threat Modelling Tool for an example product recommendation feature.}
\label{fig:data-flow-diagram} 
\end{figure}

Data Flow Diagrams (DFD) \cite{sion2018solution} are commonly used in the industry to represent data flows in complex systems. For example, Fig.~\ref{fig:data-flow-diagram} illustrates an example DFD for a single feature: product recommendations. In order to provide product recommendations, user data enters the system though a client application. More specifically, the user's name, credit card details, address and order information are sent to the purchase processing service API. In addition, the order information is sent to the service responsible for constructing customer profiles. Information related to the user's activity such as posts, sensor data or camera feeds are sent to a service producing geolocation-based inferences.

To construct a graph $G$ for this example, firstly, we create vertices $V^U$ for data types collected from the client application: user's name, credit card details, address, order information, posts, sensor data and camera feeds. Secondly, we create vertices $V^A$ for all microservices processing the data: the purchase processing service, customer profile service, geolocation inference service, etc. Then, we create a set $V^P$ containing one vertex for the purpose of this feature: serving product recommendations. Finally, we create the edges $E$ connecting vertices in $V^U$ with vertices in $V^A$ the based on the data flow, vertices in $V^A$ with other ones in $V^A$, and the vertex in $V^A$ representing the service computing product recommendations with the sole purpose vertex in $V^P$.

An example constraint on data flow of this feature could be a user disagreeing to have their order information used for serving product recommendations. Formally, we would then have a set of constraints $\mathcal{N} = \{(v_s, v_t) \: | \: v_s \in V^U, v_t \in V^P\}$ where $v_s$ is the vertex corresponding to order information and $v_t$ is our purpose vertex. To solve the problem, we would need to break the data flow from $v_s$ to $v_t$ such that the utility gained from serving product recommendations is maximised. While this example demonstrates a simple use case, in general, a data processing system may consist of many more features on which a user may impose many more constraints.

\section{Additive Model}
\label{Section: Additive Model}
As we prove in \cite{filipczuk2023consent}, in general, CDW is an $\mathcal{NP}$-hard problem. This is because in practice the valuations and utilities defined in Section~\ref{Section: Consented Data Workflow Problem} can be arbitrarily complex functions. In this section, we define a simple but practical instance of the problem, where these functions are linearly additive. In particular, we assume that the valuation function of the data going out of a vertex is linearly additive with respect to the importance of the data  on the incoming edges. That is, the value of every node's output is the sum of the values of its inputs. The linear additive model is the simplest model that captures the intuition of each input having an “added value” on the subsequent algorithm. 
This model may not apply to all settings and, e.g., a sub-additive model might be more appropriate. In many cases the linear additive model could be a reasonable approximation of a model where the inter-dependencies are difficult to measure.
Additionally, if we initialise each input edge to have an original value of 1, then the utility of a purpose node, in the this specific setting, sums up how many times the inputs ``have been used'' in algorithms before reaching the purpose node, thus giving some hint about the overall importance of an input.

In this instance of the problem we consider a DAG $G=(V,E)$ and constraints $\mathcal{N}$,  and for each edge $e = (v, v') \in E$, the valuation is defined recursively as follows:

\begin{equation}
\label{Eq: Importance LA}
\pi(e) = \sum_{e'\in in(v)} \pi(e').
\end{equation}

Similarly, we model the utility gained from processing the data for a purpose as a linearly additive function of the data valuations on the incoming edges. That is, for each reachability subgraph $G_p$, and purpose vertex $p \in V_P$, we define a utility function as follows:

\begin{equation}
\label{Eq: Utility LA}
u_p(G_p) = \sum_{e \in in(p)} \pi(e).
\end{equation}

Subsequently we present concrete heuristic implementations for the linear additive case.

\section{Algorithms}
\label{Section: Algorithms}
In this section we focus on a range of algorithms our linearly additive problem. Although there might exist multiple optimal solutions, we design our algorithms looking for a single solution $G^* = (V^*, E^*)$. While some of them offer optimal solutions, others serve as viable heuristics. Note that, even though the algorithms may not be optimal (i.e.,  utility maximising), all five algorithms always return a \emph{feasible} solution, which is a subgraph of $G$ with no path between each $(s_i, t_i) \in \mathcal{N}$ for $i \in \{1,\dots,|\mathcal{N}|\}$.

Firstly, a simple heuristic for finding a feasible solution is an algorithm that removes a random edge from each of the paths connecting $(s, t) \in \mathcal{N}$: algorithm \textsc{RemoveRandomEdge} finds all paths from $s$ to $t$ and from each of the paths selects a random edge to remove; then, before the edge is removed, the other edges whose valuation depends on the presence of the given edge in the graph must be updated. In particular, if the valuation of an edge after the update is $0$, such edge must also be removed.  This solution has a high variance but its run time  is polynomial.

Secondly, as the valuation function of the edges is additive, and because the valuation of the incoming edge of an algorithm vertex is always greater or equal than the outgoing one, the removal of the first edge of each path from $s$ to $t$ can serve as another trivial heuristic. Specifically, algorithm \textsc{RemoveFirstEdge} is very similar to \textsc{RemoveRandomEdge}, except that, instead of selecting a random edge, it removes the first edge from each path). This algorithm reflects an approach whereby the user's data is removed entirely and not even collected by the system. Similarly to \textsc{RemoveRandomEdge}, the runtime of this algorithm is polynomial.

Next, we propose a greedy algorithm running in polynomial time. This algorithm follows the heuristic of making locally optimal choices for each constraint. To do so, it uses a polynomial-time algorithm solving the Minimum Cut problem (\textsc{MinCut}) \cite{dinitz2006dinitz,edmonds1972theoretical} -- which is equivalent to the Minimum Multicut when we only have a single constraint $(s,t) \in \mathcal{N}$. Our greedy algorithm \textsc{RemoveMinCuts}, seen in Algorithm \ref{Algorithm: RemoveMinCuts},  first initialises the weights $w(e) = \pi(e) \sum_{p \in r(v)} w_p$ for all edges $e \in E$ (in lines 1 - 4), and then repeatedly calls the MinCut to find the minimum cut that solves \textsc{MinCut} for vertices $s$ and $t$ in $\mathcal{N}$ with weights $w$. For each edge in the minimum cut, it uses the updateDependencies function to update the valuations of the consecutive edges before removing the given edge. Given that \textsc{MinCut} is known to be solvable in polynomial time \cite{dinitz2006dinitz}, the outcome of this heuristic can also be found in polynomial time.

\begin{algorithm}
\caption{\textsc{RemoveMinCuts}}
\label{Algorithm: RemoveMinCuts}
\begin{flushleft}
\hspace*{\algorithmicindent} \textbf{Input:} A graph $G=(V, E)$ and a set of constraints $\mathcal{N}$. \\
 \hspace*{\algorithmicindent} \textbf{Output:} A graph $G$.
\end{flushleft}
\begin{algorithmic}[1]
\State $w \leftarrow \emptyset$
\ForAll {$e \in E$}
    \State $w(e) \leftarrow \pi(e) \sum_{p \in r(v)} w_p$
\EndFor
\ForAll{$(s, t) \in \mathcal{N}$}
    \ForAll {$e \in \textsc{MinCut}(G, w, s, t)$}
        \If {$\texttt{hasEdge}(G, e)$}
            \State $\texttt{updateDependencies}(G, e)$
            \State $\texttt{removeEdge}(G, e)$
        \EndIf
    \EndFor
\EndFor
\end{algorithmic}
\end{algorithm}

Another way of approximating the solution is by converting our problem to \textsc{MinMC}. That is, we can solve \textsc{MinMC} with weights $w(e) = \pi(e) \sum_{p \in r(v)} w_p$ for all edges $e \in E$ and then use the \textsc{MinMC} solution to find a solution. In the same way as  \textsc{RemoveMinCuts}, shown in Algorithm \ref{Algorithm: RemoveMinMC}, \textsc{RemoveMinMC} starts from initialising the weights $w$. Then, it finds the minimum multicut of graph $G$ for constraints $\mathcal{N}$. Subsequently, for each edge in the minimum multicut, it uses the updateDependencies function to update the valuations of the consecutive edges (in line 8) before removing the given edge.  

\begin{algorithm}
\caption{\textsc{RemoveMinMC}}
\label{Algorithm: RemoveMinMC}
\begin{flushleft}
\hspace*{\algorithmicindent} \textbf{Input:} A graph $G=(V, E)$, a set of constraints $\mathcal{N}$.\\
 \hspace*{\algorithmicindent} \textbf{Output:} A graph $G$.
\end{flushleft}
\begin{algorithmic}[1]
\State $w \leftarrow \emptyset$
\ForAll {$e \in E$}
    \State $w(e) \leftarrow \pi(e) \sum_{p \in r(v)} w_p$
\EndFor
\State $\texttt{multicut} \leftarrow \textsc{MinMC}(G, \mathcal{N}, w)$
\ForAll {$e \in \texttt{multicut}$}
    \If {$\texttt{hasEdge}(G, e)$}
        \State $\texttt{updateDependencies}(G, e)$
        \State $\texttt{removeEdge}(G, e)$
    \EndIf
\EndFor
\end{algorithmic}
\end{algorithm}

Finally, we propose an algorithm that can guarantee achieving an optimal solution. In Algorithm  \ref{Algorithm: BruteForce}, \textsc{BruteForce} is an exhaustive search algorithm that enumerates all feasible candidates for the solution and compares them to eventually output the one that maximises the utility. More specifically, \textsc{BruteForce} starts from finding the set of all paths $\mathcal{A}$ from $s$ to $t$ for all $(s, t) \in \mathcal{N}$, which need to be broken. In order to list all feasible multicuts of $G$ for the given $\mathcal{N}$, the Cartesian product of $\mathcal{A}$ is computed (in line 5). Then, the algorithm systematically checks the utility of $G$ after the removal of each multicut. Importantly, at the beginning of the multicut check,  copies are made of the valuation values $\pi'$ of each edge and the number of paths $p'$ the edge belongs to in $G$, as well as of the graph $G$ itself. Before an edge of the feasible multicut is removed from the copy of $G$, the valuation $\pi'$ and the number of paths $p'$ in the copy of $G$ are updated for its dependencies. At the end of the multicut check, the utility of the copy of $G$ is compared to the utility of the most optimal solution found so far. This way the algorithm can guarantee finding the optimal solution but is exponential even in the best case.

Notably, all of the presented algorithms generalize beyond the linearly additive model. What is specific to this particular model in \textsc{RemoveMinCuts}, \textsc{RemoveMinMC} and \textsc{BruteForce} is the way the weights $w(e)$ are assigned and the dependencies are updated, which depends on the valuation function $\pi(e)$ and weights $w_p$.

\begin{algorithm}
\caption{\textsc{BruteForce} }
\label{Algorithm: BruteForce}
\begin{flushleft}
\hspace*{\algorithmicindent} \textbf{Input:} A graph $G = (V, E)$ and a set of constraints $\mathcal{N}$. \\
\hspace*{\algorithmicindent} \textbf{Output:} A graph $G^*$.
\end{flushleft}
\begin{algorithmic}[1]
\State $\mathcal{A} \leftarrow \emptyset$
\ForAll {$(s, t) \in \mathcal{N}$}
    \State $\mathcal{A} \leftarrow \mathcal{A} \; \cup \; \texttt{getAllEdgePaths(}G, s, t\texttt{)}$
\EndFor

\State $\texttt{multicuts} \leftarrow \texttt{cartesianProduct}(\mathcal{A})$
\State $\texttt{maxUtility} \leftarrow 0$
\State $G^* \leftarrow G$
\ForAll {$\texttt{multicut} \in \texttt{multicuts}$}
    \State $G' \leftarrow G$
    \State $\pi', p \leftarrow \emptyset, \emptyset$
    \ForAll {$e \in E$}
        \State $\pi'(e) \leftarrow \pi(e)$
        \State $p(e) \leftarrow \sum_{p \in r(v)} w_p$
    \EndFor
    \ForAll {$e \in \texttt{multicut}$}
        \If {$\texttt{hasEdge}(G', e)$}
            \State $\texttt{updateDependencies}(G', e, \pi', p)$
            \State $\texttt{removeEdge}(G', e)$
        \EndIf
    \EndFor
    \State $\texttt{utility} \leftarrow U(G')$
    \If{$\texttt{utility} > \texttt{maxUtility}$}
        \State $\texttt{maxUtility} \leftarrow \texttt{utility}$
        \State $G^* \leftarrow G'$
    \EndIf
\EndFor
\State \Return $G^*$
\end{algorithmic}
\end{algorithm}

\section{Optimality of Solutions}
\label{Section: Properties}
Out of the five algorithms, only \textsc{BruteForce} guarantees an optimal solution. In contrast, it is clear that \textsc{RemoveRandomEdge} does not guarantee an optimal solution -- we use it as a benchmark for our evaluation. Let us analyse the properties of the solutions returned by our three remaining heuristics and prove that none of them can guarantee finding an optimal solution even for the linear setting.

Firstly, we show that a simple removal of the first edge of each path by the \textsc{RemoveFirstEdge} does not guarantee an optimal solution, i.e., for each $(s_i, t_i) \in \mathcal{N}$, there is at least one path
$P = (V_P, E_P) \in \mathcal{A}$ of the form $V_P = \{v_1,v_2,\dots,v_k\}$, $E_P = \{(v_1, v_2), (v_2, v_3), \dots, (v_{k-1}, v_k)\}$ where $v_1 = s_i$ and $v_k = t_i$. From each such path $P \in \mathcal{A}$, we could remove edge $(v_1, v_2)$. We refer to $(v_1, v_2)$ as the \textit{first edge}. Let $G$ be a data processing model where $V^U = \{ v_1 \}$, $V^A = \{ v_2 \}$, $V^P = \{ v_3, v_4 \}$, $E = \{ (v_1, v_2), (v_2, v_3), (v_2, v_4) \}$ and for each $p \in V^P$, $w_p = 1$. In addition, assume that for edge $e_1=(v_1, v_2)$, $\pi(e_1) = a$ where $a \in \mathbb{R}^+_0$, and that $\mathcal{N} = \{ (v_1, v_3) \}$. 

In such case, we use Eq.~\ref{Eq: Importance LA} to calculate the valuation of edges $e_2=(v_2, v_3)$ and $e_3 = (v_2, v_4)$, which is $\pi(e_2) = \pi(e_3) = a$. We also use Eq.~\ref{eq:global-utility-function} to calculate the initial utility of $G$, which is $U(G) = 2a$. Given that $\mathcal{N} = \{ (v_1, v_3) \}$, we establish that there is one path that needs to be disconnected in order to satisfy the constraints, i.e. $\mathcal{A} = \{ P \}$ where $P = (\{ v_1, v_2, v_3 \}, \{(v_1, v_2), (v_2, v_3)\})$.

Then, we remove the first edge $(v_1, v_2)$ from $P$. The utility of the resulting graph $G'_1$ is $U(G'_1) = 0$, since purpose vertices $v_3$ and $v_4$ are now not linked to any user vertex. However, if we instead removed the edge $(v_2, v_3)$, vertex $v_4$ would still be linked to $v_1$ and therefore the utility of the resulting graph $G'_2$ would be $U(G'_2) = a$. Thus, the removal of the first edge does not provide us with an optimal solution.

\begin{figure}
    \centering
    \begin{tikzpicture}
        \draw[fill=black] (0,1) circle (5pt);
        \draw[fill=black] (0,-1) circle (5pt);
        \draw[fill=black] (1.25,0) circle (5pt);
        \draw[fill=black] (2.5,1) circle (5pt);
        \draw[fill=black] (2.5,-1) circle (5pt);
        % nodes
        \node (A) at (0,1) {};
        \node (E) at (0,-1) {};
        \node (D) at (1.25,0) {};
        \node (B) at (2.5,1) {};
        \node (C) at (2.5,-1) {};
        \path [->,-{Stealth[scale=1.5]}]   (A) edge node[midway,above] {$a$} (D); 
        \path [->,-{Stealth[scale=1.5]}]   (E) edge node[midway,above] {$b$} (D); 
        \path [->,-{Stealth[scale=1.5]}]   (D) edge node[midway,above,rotate=40,text width=2cm,align=center] {$a+b$} (B); 
        \path [->,-{Stealth[scale=1.5]}]   (D) edge node[midway,above,rotate=-40] {$a+b$} (C);
        \node (L) at (-0.5,1) [text width=3cm,align=center] {$s_1$};
        \node (L) at (-0.5,-1) [text width=3cm,align=center] {$s_2$};
        \node (L) at (1.25,0.5) [text width=3cm,align=center] {$v_1$};
        \node (L) at (3,1) [text width=3cm,align=center] {$t_1$};
        \node (P) at (3,-1) [text width=3cm,align=center] {$t_2$};
    \end{tikzpicture}
    \caption{A data processing model where $V^U = \{ s_1, s_2 \}$, $V^A = \{ v_1 \}$, $V^P = \{ t_1, t_2 \}$ and $E = \{ (s_1, v_1), (s_2, v_1), (v_1, t_1), (v_1, t_2) \}$.}
    \label{Fig: Case2}
\end{figure}
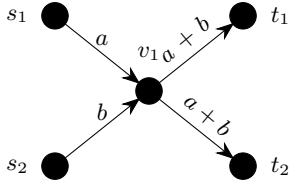

Furthermore, \textsc{RemoveMinCuts} does not lead to an optimal solution. Consider a series of graphs $G^0,G^1,\dots,G^{|\mathcal{N}|}$. These graphs are computed recursively such that $G^0 = G = (V, E)$ and for all $i \in \{1,\dots,|\mathcal{N}|\}$, $G^i = (V, E^i)$ where $E^i = E^{i-1} \setminus \textsc{MinCut}(G^{i - 1}, s_i, t_i, w)$ corresponds to $i$-th pair of vertices in $\mathcal{N}$. Given a solution to \textsc{MinCut}, one could transform the problem into a repeated \textsc{MinCut} problem looking for $G^{|\mathcal{N}|}$. While this approach leads to a feasible solution, we show that $G^{|\mathcal{N}|}$ is not an optimal solution. Let $G$ be a graph where $V^U = \{ s_1, s_2 \}$, $V^A = \{ v_1 \}$, $V^P = \{ t_1, t_2 \}$, $E = \{ (s_1, v_1), (s_2, v_1), (v_1, t_1), (v_1, t_2) \}$  and for each $p \in V^P$, $w_p = 1$. Assume that for $e_1=(s_1, v_1)$, $\pi(e_1) = a$ and for $e_2=(s_2, v_1)$, $\pi(e_2) = b$, where $a,b \in \mathbb{R}^+_0$ and $a > b$ (see Fig.~\ref{Fig: Case2}). In addition, $\mathcal{N} = \{ (s_1, t_1), (s_1, t_2) \}$. We look for $G^2 = (V, E \setminus \textsc{MinCut}(G^1, s_1, t_1, w))$. Thus, we first calculate $G^1 = (V, E \setminus \textsc{MinCut}(G, s_1, t_1, w))$. We observe that there is one path $P = (\{s_1, v_1, t_1\}, \{(s_1, v_1), (v_1, t_1) \}) $ between vertices $s_1$ and $t_1$. Because $a > b$, $w((v_1, t_1)) = a + b < w((s_1, v_1)) = 2a$. Thus, $G^1 = (V, E \setminus \{(v_1, t_1)\})$. With this information, we return to looking for $G^2$. We observe that there is one path $P = (\{s_1, v_1, t_2\}, \{(s_1, v_1), (v_1, t_2) \}) $ between vertices $s_1$ and $t_2$. However, now $w((s_1, v_1)) = a$ and $w((v_1, t_2)) = a + b$. So, $w((s_1, v_1)) < w((v_1, t_2))$ and $G^2 = (V, E \setminus \{(s_1, v_1), (v_1, t_1)\})$. After that, we calculate $U(G^2) = b$. However, we can see that to obtain an optimal solution, it is sufficient to remove $(s_1, v_1)$ only: the optimal solution is $G^* = (V, E \setminus \{(s_1, v_1)\})$, because its utility is $U(G^*) = 2b$. Thus, $G^2$ is not an optimal solution.

Intuitively, it is reasonable to assume that the optimal solution requires removing no more than one edge per path between $s$ and $t$ for each  $(s,t) \in  \mathcal{N}$. Let $T \subseteq V^P$ be a set of purpose vertices such that for all $(s, t) \in \mathcal{N}$, $t \in T$. If for each path $P=(V_P, E_P) \in \mathcal{A}$ there is exactly one edge $e \in E_P$ such that $e \in E_{MinMC}$, then the removal of  $E_{MinMC}$ reduces the utility of a purpose vertex $t \in T$ by $\sum_{e \in E_t} \pi(e)$ where $E_t \subseteq E_{MinMC}$ is a set of those edges in $E_{MinMC}$ that are within the reachability subgraph of $t$, $G_t$. Thus, if the resulting graph after the removal of set $E_{MinMC}$ from $G$ is $G' = (V, E')$, then using Eq.~\ref{eq:global-utility-function}, the total loss of utility is:

\begin{equation}
\label{Eq: properties-1}
    U(G) - U(G') = \sum_{t \in T} \sum_{e \in E_t} w_t \pi(e).
\end{equation}

This is equivalent to the following equation:
\begin{equation}
\label{Eq: label-inside-the-proof}
    U(G') = U(G) - \sum_{e \in E \setminus E'} \pi(e) \sum_{t \in T} w_t.
\end{equation}

In fact, if we plug Eq.~\ref{Eq: label-inside-the-proof} into Equation \ref{eq:optimisation_problem}, we have:

\begin{equation}
    G^* = arg \: \max_{G'} \: \{ U(G) - \sum_{e \in E \setminus E'} \pi(e) \sum_{t \in T} w_t \}.
\end{equation}

Equivalently, we are looking for a subgraph where:

\begin{equation}
\label{Eq: label2-inside-the-proof}
    E^* = E  \: \setminus  \: \{ arg \: \min_{E \setminus E'} \: \sum_{e \in E \setminus E'} \pi(e) \sum_{t \in T} w_t \}.
\end{equation}

Thus, $E^*$ is the set difference of $E$ and the minimum multicut of $G$ given $\mathcal{N}$, where the edge weight is $w(e) = \pi(e) \sum_{t \in T} w_t$. In more detail, the minimum multicut with edge weights $w(e) = \pi(e) \sum_{t \in T} w_t$ can be expressed as: 
\begin{equation}
\label{Eq: label3-inside-the-proof}
    E_{MinMC} = arg \: \min_{E'} \: \sum_{e \in E'} \pi(e) \sum_{t \in T} w_t.
\end{equation}

If we plug Eq.~\ref{Eq: label2-inside-the-proof} into Eq.~\ref{Eq: label3-inside-the-proof}, then what we are looking for is $G^*=(V, E^*)$ where:

\begin{equation}
    E^* = E \: \setminus \: E_{MinMC}.
\end{equation}

Since $E_{MinMC}$ is a solution to \textsc{MinMC}, there is $G^* = (V, E \setminus E_{MinMC})$.

However, removing a single edge from each path does not always result in an optimal solution. Consider a graph $G$ where $V^U = \{ s_1, s_2 \}$, $V^A = \{ v_1 \}$, $V^P = \{ t_1, t_2 \}$, $E = \{ (s_1, v_1), (s_2, v_1), (v_1, t_1), (v_1, t_2) \}$ and for each $p \in V^P$, $w_p = 1$. In addition, assume that for $e_1=(s_1, v_1)$, $\pi(e_1) = a$ and for $e_2=(s_2, v_1)$, $\pi(e_2) = b$, where $a,b \in \mathbb{R}^+_0$ and $a > b$ (see Fig.~\ref{Fig: Case2}). Now, let the set of constraints be as follows: $\mathcal{N} = \{ (s_1, t_1), (s_1, t_2), (s_2, t_1) \}$. We can see that the optimal solution in this case is $G^* = (V, \{ (s_2, v_1), (v_1, t_2) \})$. However, since $(s_1, t_1) \in \mathcal{N}$ and there is a path from $s_1$ to $t_1$ in the original graph $G$, we can observe that the optimal solution $G^*$ does not contain two edges $(s_1, v_1)$ and $(v_1, t_2)$ from that path.Therefore, in general, it is not true that \textsc{RemoveMinMC} can guarantee finding an optimal solution.

\section{Experimental Evaluation}
\label{Section: Evaluation}
To evaluate the algorithms' performance empirically, we perform experiments on the University of Southampton High Performance Computing service \textit{Iridis 4}\footnote{The Iridis Compute Cluster, \url{https://cmg.soton.ac.uk/iridis}.}, measuring their performance on synthetic data. Our graph generation method includes the following parameters:
\begin{itemize}
    \item number of constraints $|\mathcal{N}|$;
    \item number of vertices $|V|$;
    \item path length $k$ -- for any $(s, t) \in \mathcal{N}$, if there is a path $P = 
    \{(v_1, v_2)\dots(v_{k-1},v_k))\}$  such that $v_1 = s$ and $v_k = t$, then $k$ defines the number of workflow stages, i.e. data that `flows' from $s$ to $t$ though $k - 2$ algorithm nodes;
    \item vertex distribution vector $X_k$ -- proportions of vertices at different data flow stages, e.g. a setting $X_k = (50\%, 25\%, 10\%, 10\%, 5\%)$ represents a scenario for $k = 5$ where half of the vertices are the user data vertices, 45\% are the algorithm vertices and 5\% are the number of the purpose vertices (NU - non-uniform distribution, U - uniform distribution); 
    \item minimum density $d$ -- the proportion of initially generated edges between any two workflow stages.
\end{itemize}

\begin{table}
\caption{Parameter configurations for datasets 1, 2 and 3.}
\label{Tab:parameters}
\begin{tabular}{c|c|c|c|c|c|}
\cline{2-6}
\multirow{2}{*}{} & \multicolumn{3}{c|}{\textbf{Dataset 1}} & \multirow{2}{*}{\textbf{Dataset 2}} & \multirow{2}{*}{\textbf{Dataset 3}} \\ \cline{2-4}
  & \textbf{a}     & \textbf{b}     & \textbf{c}    &                     &                     \\ \hline\hline
\multicolumn{1}{|c||}{\textbf{$|\mathcal{N}|$}} &   1 -- 50     &   1 -- 50  & 1 -- 50  &    10   &              5                    \\ \hline
\multicolumn{1}{|c||}{\textbf{$|V|$}} &  100      &  1000      &  100     &       150 -- 5000              &   100 -- 10000                  \\ \hline
\multicolumn{1}{|c||}{\textbf{$k$}} &  5      &    5    &   5    &   3 -- 50                  &      5               \\ \hline
\multicolumn{1}{|c||}{\textbf{$X_k$}} &   NU     &    NU    &    U   &    U                 &      NU               \\ \hline
\multicolumn{1}{|c||}{\textbf{$d$}} &   0     &     0   &  20\%     &  0                   &     0                \\ \hline
\end{tabular}
\end{table}

We prepare three datasets with different configurations of the above parameters, specified in Tab.~\ref{Tab:parameters}.

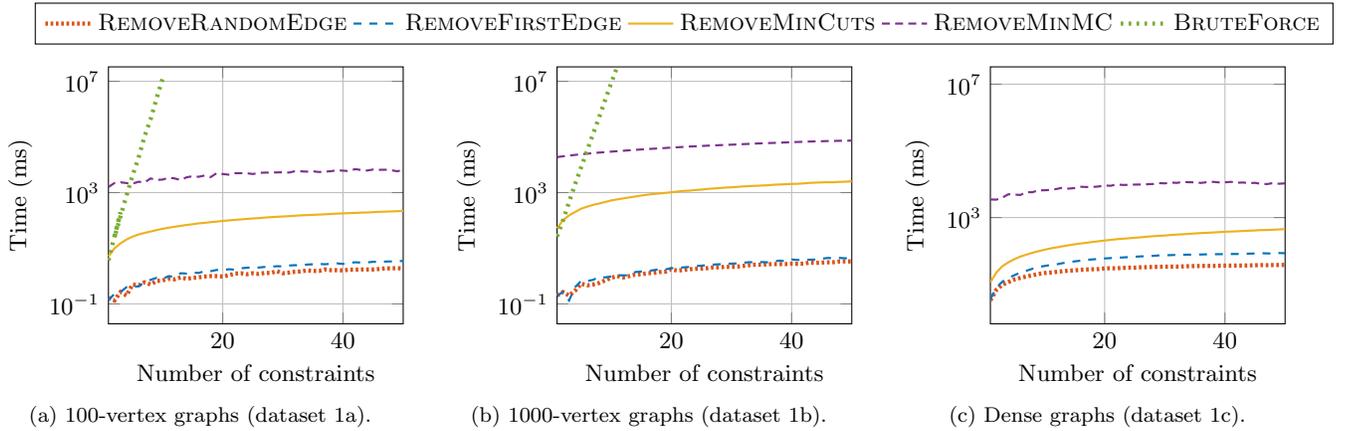
\begin{figure*}[t!]
\centering
\begin{subfigure}{.3\textwidth}
\begin{tikzpicture}
    \begin{semilogyaxis}[
        height=5cm,
        width=5.5cm,
        grid=major,
        legend style={at={(-0.25,1.25)}, anchor=north west,legend columns = 1},
        legend columns=5,
        xmin=1,
        xmax=50,
        ymax=33020228,
        yticklabel style={/pgf/number format/fixed},
        xlabel=Number of constraints,
        ylabel=Time (ms)]
            \addlegendentry{\textsc{RemoveRandomEdge}}
            \csvdenselydotted{matlab_color_2}{./raw_data/a2_sparse_cons_100_times.csv}
            \addlegendentry{\textsc{RemoveFirstEdge}}
            \csvdashed{matlab_color_1}{./raw_data/a1_sparse_cons_100_times.csv}
            \addlegendentry{\textsc{RemoveMinCuts}}
            \csvline{matlab_color_3}{./raw_data/a3_sparse_cons_100_times.csv}
            \addlegendentry{\textsc{RemoveMinMC}}
            \csvdenselydashed{matlab_color_4}{./raw_data/a4_sparse_cons_100_times.csv}
            \addlegendentry{\textsc{BruteForce}}
            \csvdotted{matlab_color_5}{./raw_data/a7_sparse_cons_100_times.csv}
    \end{semilogyaxis}
\end{tikzpicture}
\caption{100-vertex graphs (dataset 1a).}
\label{fig:100v_constraints_vs_time}
\end{subfigure}\qquad
\begin{subfigure}{.3\textwidth}
\vspace*{0.8cm}
\begin{tikzpicture}
    \begin{semilogyaxis}[
        height=5cm,
        width=5.5cm,
        grid=major,
        xmin=1,
        xmax=50,
        ymax=33020228,
        yticklabel style={/pgf/number format/fixed},
        xlabel=Number of constraints,
        ylabel=Time (ms)]
            \csvdenselydotted{matlab_color_2}{./raw_data/a2_sparse_cons_times.csv}
            \csvdashed{matlab_color_1}{./raw_data/a1_sparse_cons_times.csv}
            \csvline{matlab_color_3}{./raw_data/a3_sparse_cons_times.csv}
            \csvdenselydashed{matlab_color_4}{./raw_data/a4_sparse_cons_times.csv}
            \csvdotted{matlab_color_5}{./raw_data/a5_1b_times.csv}
    \end{semilogyaxis}
\end{tikzpicture}
\caption{1000-vertex graphs (dataset 1b).}
\label{fig:1000v_constraints_vs_time}
\end{subfigure}\qquad
\begin{subfigure}{.3\textwidth}
\vspace*{0.8cm}
\begin{tikzpicture}
    \begin{semilogyaxis}[
        height=5cm,
        width=5.5cm,
        grid=major,
        xmin=1,
        xmax=50,
        ymax=33020228,
        ytick={0.1,10^3,10^7},
        yticklabel style={/pgf/number format/fixed},
        xlabel=Number of constraints,
        ylabel=Time (ms)]
            \csvdenselydotted{matlab_color_2}{./raw_data/a2_dense_cons_100_times.csv}
            \csvdashed{matlab_color_1}{./raw_data/a1_dense_cons_100_times.csv}
            \csvline{matlab_color_3}{./raw_data/a3_dense_cons_100_times.csv}
            \csvdenselydashed{matlab_color_4}{./raw_data/a4_dense_cons_100_times.csv}
    \end{semilogyaxis}
\end{tikzpicture}
\caption{Dense graphs (dataset 1c).}
\label{fig:100v_d_constraints_vs_time}
\end{subfigure}
\caption{The number of constraints vs. the runtime of the algorithms in graphs from dataset 1.}
\label{Fig: dataset1}
\end{figure*}

We measure the runtime of the algorithms on 100-vertex graphs (dataset 1a), on 10 times larger, 1000-vertex graphs (dataset 1b) and on 100-vertex graphs where the number of edges between each level of data processing is at least 20\% of all possible edges (dataset 1c). We observe that the runtime of \textsc{BruteForce} increases rapidly with the increasing number of constraints, reaching an average time of 14838508.46 ms (i.e. over 4 h) for just 10 constraints on dataset 1a and 8563968 ms (i.e. over 2 h) on dataset 1b. For dataset 1c, in most cases, \textsc{BruteForce} is  unable to return a result in 60 hours even for just a single pair of constraints. Thus, although \textsc{BruteForce} guarantees finding an optimal solution, its average runtime makes this algorithm impractical. At the same time, the solver-based \textsc{RemoveMinMC} can reach an approximate solution for even 50 constraints on average in 6.51s on dataset 1a, 74.1s on dataset 1b and 11s on dataset 1c. \textsc{RemoveMinCuts} can on average find an approximate solution for 50 constraints in 220.2 ms on dataset 1a, 2.55s on dataset 1b and 450.57 ms on dataset 1c.

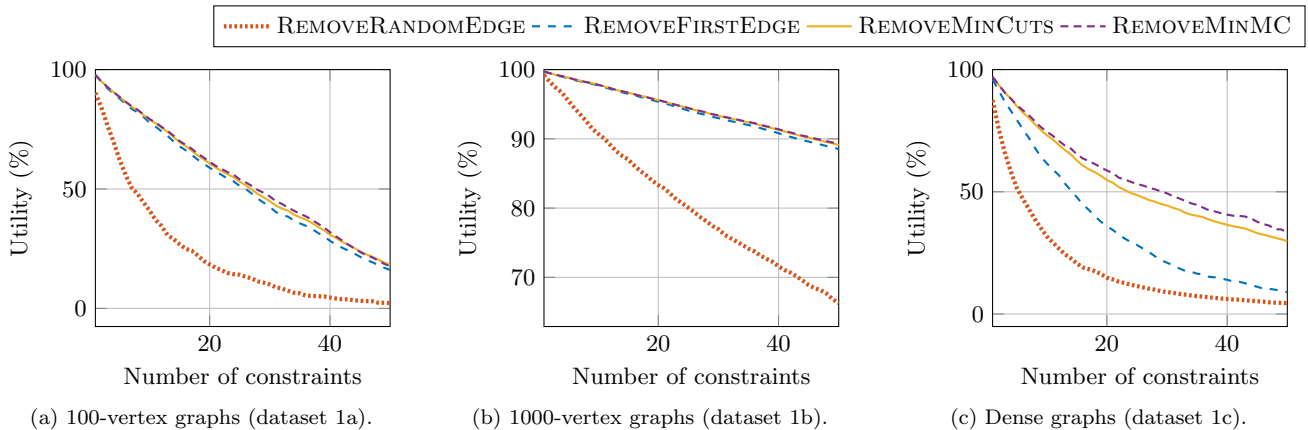
\begin{figure*}
\centering
\begin{subfigure}{.3\textwidth}
\begin{tikzpicture}
    \begin{axis}[
        height=5cm,
        width=5.5cm,
        grid=major,
        legend style={at={(0.4,1.25)}, anchor=north west,legend columns = 1},
        legend columns=5,
        xmin=1,
        xmax=50,
        ymax=100,
        yticklabel style={/pgf/number format/fixed},
        xlabel=Number of constraints,
        ylabel=Utility (\%)]
            \addlegendentry{\textsc{RemoveRandomEdge}}
            \csvdenselydotted{matlab_color_2}{./raw_data/a2_sparse_cons_100_utilities.csv}
            \addlegendentry{\textsc{RemoveFirstEdge}}
            \csvdashed{matlab_color_1}{./raw_data/a1_sparse_cons_100_utilities.csv}
            \addlegendentry{\textsc{RemoveMinCuts}}
            \csvline{matlab_color_3}{./raw_data/a3_sparse_cons_100_utilities.csv}
            \addlegendentry{\textsc{RemoveMinMC}}
            \csvdenselydashed{matlab_color_4}{./raw_data/a4_sparse_cons_100_utilities.csv}
    \end{axis}
\end{tikzpicture}
\caption{100-vertex graphs (dataset 1a).}
\label{fig:100v_constraints_vs_utility}
\end{subfigure}\qquad
\begin{subfigure}{.3\textwidth}
\vspace*{0.8cm}
\begin{tikzpicture}
    \begin{axis}[
        height=5cm,
        width=5.5cm,
        grid=major,
        xmin=1,
        xmax=50,
        ymax=100,
        yticklabel style={/pgf/number format/fixed},
        xlabel=Number of constraints,
        ylabel=Utility (\%)]
            \csvdenselydotted{matlab_color_2}{./raw_data/a2_sparse_cons_utilities.csv}
            \csvdashed{matlab_color_1}{./raw_data/a1_sparse_cons_utilities.csv}
            \csvline{matlab_color_3}{./raw_data/a3_sparse_cons_utilities.csv}
            \csvdenselydashed{matlab_color_4}{./raw_data/a4_sparse_cons_utilities.csv}
    \end{axis}
\end{tikzpicture}
\caption{1000-vertex graphs (dataset 1b).}
\label{fig:1000v_constraints_vs_utility}
\end{subfigure}\qquad
\begin{subfigure}{.3\textwidth}
\vspace*{0.8cm}
\begin{tikzpicture}
    \begin{axis}[
        height=5cm,
        width=5.5cm,
        grid=major,
        xmin=1,
        xmax=50,
        ymax=100,
        yticklabel style={/pgf/number format/fixed},
        xlabel=Number of constraints,
        ylabel=Utility (\%)]
            \csvdenselydotted{matlab_color_2}{./raw_data/a2_dense_cons_100_utilities.csv}
            \csvdashed{matlab_color_1}{./raw_data/a1_dense_cons_100_utilities.csv}
            \csvline{matlab_color_3}{./raw_data/a3_dense_cons_100_utilities.csv}
            \csvdenselydashed{matlab_color_4}{./raw_data/a4_dense_cons_100_utilities.csv}
    \end{axis}
\end{tikzpicture}
\caption{Dense graphs (dataset 1c).}
\label{fig:100v_d_constraints_vs_utility}
\end{subfigure}
\caption{The number of constraints vs. graph utility after applying the algorithms on graphs from dataset 1.}
\label{fig:constraints_vs_utility}
\end{figure*}

\begin{table}
\caption{Comparison of the graph's utility after applying \textsc{RemoveMinMC} and \textsc{BruteForce}.}
\label{tab:utility-comparison}
\begin{tabular}{|c|c|c|c|c|}
\hline
Number of & \multicolumn{2}{c|}{\textsc{RemoveMinMC}} & \multicolumn{2}{c|}{\textsc{BruteForce}} \\ \cline{2-5} 
constraints & \% of original & SE & \% of original & SE \\ \hline

1                     &  97.79             &   0.32            &  97.79              &    0.32            \\
2                     &  95.08             &   0.35            &  95.08              &    0.35            \\
3                     &  92.71             &  0.58             &  92.71              &   0.58             \\
4                     &  90.62             &  0.57             &  90.63              &    0.57            \\
5                     &  88.59             &  0.75             &   88.65             &    0.75            \\
6                     &  86.59             &  0.72             &   86.66             &    0.72            \\
7                     &  84.71             &   0.72            &   84.77             &    0.71            \\
8                     &  83.22             &  0.70             &   83.28             &    0.70            \\
9                     & 81.24              &  0.69             & 81.33                &  0.69              \\
10                    &  79.30             &  0.69             &  79.39              &    0.68            \\ \hline
\end{tabular}
\end{table}

Moreover, we observe a change in the graph's utility as the number of constraints grows. Although the exponential runtime of the \textsc{BruteForce} algorithm means we cannot run the experiments for more than 10 constraints, we still compare the results to \textsc{RemoveMinMC} for this limited setting. Results are presented in Tab.\ref{tab:utility-comparison} and show that the utility using \textsc{RemoveMinMC} is nearly optimal in this case. Although the algorithm is only guaranteed to be optimal for specific settings, this empirical outcome suggests that, for graphs with a relatively small number of constraints, \textsc{RemoveMinMC} is likely to provide very accurate solutions. Moreover, Fig.~\ref{fig:constraints_vs_utility}, shows that the differences in utility between algorithms is more evident when the graphs are denser, resulting in significantly poorer performance of \textsc{RemoveMinCuts}, \textsc{RemoveFirstEdge} and \textsc{RemoveRandomEdge}. Nonetheless, \textsc{RemoveMinMC} provides the best solution for all three types of graphs.

\begin{figure}[t!]
    \centering
    \begin{subfigure}{.4\textwidth}
    \begin{tikzpicture}
    \begin{axis}[
        height=5cm,
        width=7.5cm,
        grid=major,
        legend style={at={(-0.01,1.15)}, anchor=south west,legend columns = 1},
        legend columns=2,
        xmin=1,
        xmax=850,
        yticklabel style={/pgf/number format/fixed},
        xlabel=Number of paths,
        ylabel=Time (ms)]
            \csvpointstwo{matlab_color_2}{./raw_data/a2_dense_cons_100_scatter_times.csv}
            \addlegendentry{\textsc{RemoveRandomEdge}}
            \csvpointstwo{matlab_color_1}{./raw_data/a1_dense_cons_100_scatter_times.csv}
            \addlegendentry{\textsc{RemoveFirstEdge}}
            \csvpointstwo{matlab_color_3}{./raw_data/a3_dense_cons_100_scatter_times.csv}
            \addlegendentry{\textsc{RemoveMinCuts}}
            \csvpointstwo{matlab_color_4}{./raw_data/a4_dense_cons_100_scatter_times.csv}
            \addlegendentry{\textsc{RemoveMinMCs}}
    \end{axis}
\end{tikzpicture}
\end{subfigure}
\begin{subfigure}{.42\textwidth}
    \includegraphics[width=\textwidth]{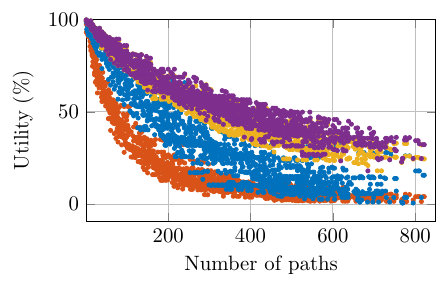}
\end{subfigure}
\caption{No. of paths vs. runtime and utility (dataset 1c).}
\label{fig:paths_vs_utility}
\end{figure}

Next, we observe how the execution time depends on the number of paths between pairs of vertices that connect the constraints. 
Fig.~\ref{fig:paths_vs_utility} presents a scatter plot of the runtime of the algorithms and distribution of utility for dataset 1c. In particular, we can see that, in case of \textsc{RemoveMinCuts} and \textsc{RemoveMinMC}, the runtimes increase almost linearly with respect to the number of paths. However, the execution times for these two algorithms differ significantly. For example, for a graph where 822 paths are required to be broken, \textsc{RemoveMinMC} takes 12116 ms to return a solution, whereas \textsc{RemoveMinCuts} can provide one in only 472 ms. Similarly, we can see that the utility  decreases as the number of paths connecting constraints increases. Yet again, the utility after executing \textsc{RemoveMinMC} tends to decrease the slowest, reaching on average the utility of 32.27\% of the original utility for the graph with 822 paths to be broken. For the same graph, the next best solution is \textsc{RemoveMinCuts} with an accuracy of 24.58\%. For comparison, \textsc{RemoveFirstEdge} achieves on average utility of 15.73\%. 

\begin{figure}
    \centering
    \begin{tikzpicture}
    \begin{semilogyaxis}[
        height=5cm,
        width=7.5cm,
        grid=major,
        legend pos=south east,
        xmin=2,
        xmax=50,
        ymax=14838508,
        yticklabel style={/pgf/number format/fixed},
        xlabel=Path length,
        ylabel=Time (ms)]
         \csvdenselydotted{matlab_color_2}{./raw_data/a2_path_length__times.csv}
         \addlegendentry{\textsc{RemoveRandomEdge}}
         \csvline{matlab_color_1}{./raw_data/a1_path_length__times.csv}
         \addlegendentry{\textsc{RemoveFirstEdge}}
         \csvdenselydashed{matlab_color_3}{./raw_data/a3_path_length__times.csv}
         \addlegendentry{\textsc{RemoveMinCuts}}
         \csvdotted{matlab_color_4}{./raw_data/a4_path_length__times.csv}
         \addlegendentry{\textsc{RemoveMinMC}}
         \csvline{matlab_color_5}{./raw_data/a5_path_length__times.csv}
    \end{semilogyaxis}
\end{tikzpicture}
    \caption{Path length vs. time in sparse graphs (dataset 2).}
    \label{fig:path_length_vs_time}
\end{figure}
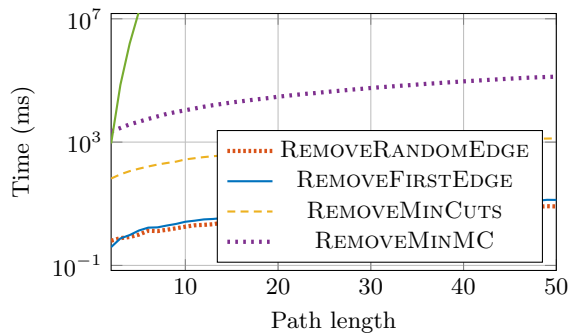

Next, we apply the algorithms to graphs with a constant number of paths. Since only edges connected to purpose vertices affect the utility, increasing the lengths of the paths on its own does not affect the utility. Thus, we focus on the algorithm runtime as the path length grows. In Fig.~\ref{fig:path_length_vs_time}, we consider sparse graphs with vertices distributed uniformly with the same number of user data vertices as purpose vertices (dataset 2). We can see that as the path length grows, the runtime in case of \textsc{BruteForce} increases faster.

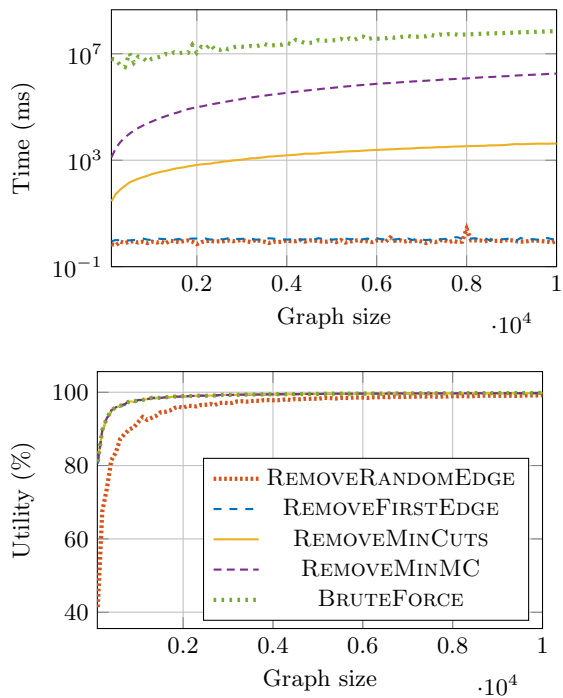
\begin{figure}
    \centering
    \begin{subfigure}{.4\textwidth}
    \centering
    \begin{tikzpicture}
    \begin{semilogyaxis}[
        height=5cm,
        width=7.5cm,
        grid=major,
        xmin=100,
        xmax=10000,
        yticklabel style={/pgf/number format/fixed},
        xlabel=Graph size,
        ylabel=Time (ms)]
         \csvdenselydotted{matlab_color_2}{./raw_data/a2_remove_random_.csv}
         \csvdashed{matlab_color_1}{./raw_data/a1_remove_first_.csv}
         \csvline{matlab_color_3}{./raw_data/a3_remove_st_cuts_.csv}
         \csvdenselydashed{matlab_color_4}{./raw_data/a4_solver_.csv}
         \csvdotted{matlab_color_5}{./raw_data/a7_bf_.csv}
    \end{semilogyaxis}
\end{tikzpicture}
\end{subfigure}
\par\bigskip
\begin{subfigure}{.4\textwidth}
\centering
    \begin{tikzpicture}
    \begin{axis}[
        height=5cm,
        width=7.5cm,
        grid=major,
        xmin=90,
        xmax=10000,
        legend pos=south east,
        yticklabel style={/pgf/number format/fixed},
        xlabel=Graph size,
        ylabel=Utility (\%)]
            \csvdenselydotted{matlab_color_2}{./raw_data/a2_utilities.csv}
            \addlegendentry{\textsc{RemoveRandomEdge}}
            \csvdashed{matlab_color_1}{./raw_data/a1_utilities.csv}
            \addlegendentry{\textsc{RemoveFirstEdge}}
            \csvline{matlab_color_3}{./raw_data/a3_utilities.csv}
            \addlegendentry{\textsc{RemoveMinCuts}}
            \csvdenselydashed{matlab_color_4}{./raw_data/a4_utilities.csv}
            \addlegendentry{\textsc{RemoveMinMC}}
            \csvdotted{matlab_color_5}{./raw_data/a7_bf_utilities.csv}
            \addlegendentry{\textsc{BruteForce}}
    \end{axis}
\end{tikzpicture}
\end{subfigure}
    \caption{Graph size vs. runtime and utility (dataset 3).}
    \label{fig:graph_size_vs_time}
\end{figure}

Lastly, we analyse how the number of vertices in the graph impacts the runtime and the utility of the graph after applying the algorithms. To do this, we run the algorithms on graphs from dataset 3. As the number of paths between the constraints and their length are equal for these graphs, in Fig.~\ref{fig:graph_size_vs_time} we can see that the size of the graph has only a slight impact on the execution time for \textsc{BruteForce}. In addition, \textsc{RemoveMinCuts} is faster on average compared to \textsc{BruteForce} and \textsc{RemoveMinMC}. Considering the utility,  Fig.~\ref{fig:graph_size_vs_time} shows that the graph size does not have significant impact on utility when the number of paths between the constraints and their length remain equal for the graphs.

\section{Open Problems}
\label{section:openproblems}

In this paper, we designed a theoretical mechanism where the data flow is structured as a graph and privacy constraints collected from the user point to pairs of vertices in this graph. While our approach is effective and finds a nearly optimal solution for large graphs, it also creates a range of open problems and challenges. 

First, for our algorithms to deliver value, a service provider needs to build an accurate graph. In particular, for large-scale systems with thousands of microservices built on billions lines of code in multiple languages, identifying data flows and all the purposes they are processed for, is challenging and time-consuming. Therefore, future work should develop automated tools to support graph generation. 

Second, some of our effective algorithms rely on the simplifying assumption that the value of different information sources is additive. While this is a reasonable starting point and can be a reasonable approximation in some settings, in practice the value may be subadditive (e.g. in the case of redundant data) or superadditive (when data complements each other). As data processing systems keep expanding, future work should focus on exploring more variable models that align the utility and valuation functions with real-world use cases.
While the algorithms proposed in this article conceptually generalize to more complex settings, models with different utility and valuation functions may open opportunities for designing more efficient algorithms. For example, in certain machine learning as well as statistical algorithms there is already work that addresses masking or distorting the input with noise. This gives rise to an exciting generalisation of our framework, where utility is affected not by removing the input edge but by distorting the input, and in the future we plan to explore this.

Third, we show that the problem in general is NP-hard. This result provides us with the lower bound on the complexity of the problem. The upper bound, however, depends largely on the complexity of the selected valuation and utility functions. In the future, we plane to investigate the upper bound of the problem in different settings. 

In addition, there are several open problems regarding the scalability of the problem. Currently, our solution needs to be recomputed every time a new user enters the system or when an existing user updates their constraints. What is more, every time a change is made, some of the algorithms that process the data would need to be re-run as well, which could be costly. At the same time, there could be many users of the same type, i.e. with similar privacy constraints, and a limited number of different user types which can be known in advance. To take advantage of this, users of the same type could, e.g., be treated as a single user to cope with thousands and even millions of users. This way, if new users enter the system, a new solution can be found quickly. Generally, as more and more users have privacy constraints, new methods are needed that take into account scalability by re-using some of the computation performed for the previous solutions, as well as the costs of making changes.

Finally, there are plenty of opportunities to consider richer types of privacy constraints and user preferences. For example, users may have constraints on combinations of different data types for a specific purpose (e.g. a user may say `\textit{I'm okay with you using my data for advertising, but don't combine my location with my purchase history}') or time restrictions on data processing (\textit{`I'm okay with you sharing my purchase history, but I don't want them to keep it for more than 30 days}'). Future work should formulate new problems around these constraints, as well as constraint the secondary re-use of data later in the data lifecycle.

\section{Related Work}
\label{Section: Related Work}
With the scale of data collection and processing growing vastly in the recent years, there has been an emergence of initiatives and tools that aim to aid users in controlling the flow of their personal data. Early efforts include the Platform for Privacy Preferences (P3P) which aimed to enable machine-readable privacy policies \cite{cranor2002web}. Such privacy policies could be automatically retrieved by Web browsers and other 
tools that can display symbols, prompt users, or take other appropriate actions. Users were able to communicate their privacy constraints to these so-called \textit{user agents} as a list of rules expressed in a P3P Preference Exchange Language (APPEL) \cite{cranor2002web}. The agents were then able to compare each policy against the user’s constraints and assist the user in deciding when to exchange data with websites \cite{cranor2002web}.

However, the P3P lacked a mechanism that would allow for enforcement of the privacy policy within the enterprise and for management of users' individual privacy preferences \cite{ashley2002privacy,agrawal2002hippocratic,kaufman2002social}. Thus, a new approach was proposed \cite{ashley2002privacy} where service providers would publish privacy policies, collect and manage user preferences and consent, and enforce the policies throughout their systems. Based on this framework, the Platform for Enterprise Privacy Practices (E-P3P) was developed \cite{ashley2002p3p,karjoth2002platform}. Our work builds on this idea and proposes how to satisfy the users' individual preferences \emph{optimally}.

Nonetheless, more general approaches that go beyond the P3P were required to address the need for policy enforcement. To that end, Hippocratic databases were proposed \cite{agrawal2002hippocratic,agrawal2003implementing,agrawal2004auditing,lefevre2004limiting,agrawal2005extending} as `database systems that take responsibility for the privacy of data they manage'. There, personal data was associated with the purposes it was collected for, and metadata such as retention period. Given privacy constraints, the so-called Privacy Constraint Validator would check if the policy is acceptable to the user. Recently, an approach to enable even more fine-grained control of such policies within relational databases was presented~\cite{konstantinidis-enabling-2021}. Here, we extend this approach.

Complementing our work, related work has proposed methods for privacy policy-compliant data processing. For example, once it is known how the user's privacy constraints can be optimally satisfied (e.g., using our proposed approach), it is possible to ensure that the datasets used by the data-processing algorithms align with these constraints \cite{debruyne2019towards, debruyne2020just}. Moreover, data processing techniques can be selected based on the consented types of the input data \cite{wang2007respecting}. Furthermore, to make sure that the policies are enforced, a system for checking data usage policies automatically at query runtime has been proposed \cite{upadhyaya2015automatic}.

\section{Acknowledgements}
E. Gerding was partially funded by the EPSRC-funded platform grant “AutoTrust: Designing a Human-Centred Trusted, Secure, Intelligent and Usable Internet of Vehicles” (EP/R029563/1). G. Konstantinidis was partially funded by the UKRI Horizon Europe guarantee funding scheme for the Horizon Europe projects RAISE (No. 101058479) and UPCAST (No. 101093216).

\bibliographystyle{abbrv} 
\bibliography{main}  
\balancecolumns
\end{document}